# Vortex dynamics in amorphous MoSi superconducting thin films

*Zhengyuan Liu[a], Bingcheng Luo[*a], Labao Zhang[b], Boyu Hou[a], Danyang Wang[c]*

[*]Corresponding author

[a]School of Physical Science and Technology, Northwestern Polytechnical University, Xi'an, Shaanxi, 710072, China

Electronic mail: luobingcheng@nwpu.edu.cn

[b]Research Institute of Superconductor Electronics, Nanjing University, Nanjing 210093, China

[c]School of Materials Science and Engineering, The University of New South Wales, Sydney, NSW2052, Australia

**ABSTRACT**

Vortex dynamics in superconductors have received a great deal of attention from both fundamental and applied researchers over the past few decades. Because of its critical role in the energy relaxation process of type-II superconductors, vortex dynamics have been deemed a key factor for the emerging superconducting devices, but the effect of irradiation on vortex dynamics remains unclear. With the support of electrical transport measurements under external magnetic fields and irradiation, photon effect on vortex dynamics in amorphous MoSi (*a*-MoSi) superconducting thin films are investigated in this work. The magnetic-field-dependent critical vortex velocity $v^*$ derived from the Larkin–Ovchinnikov model is not significantly affected by irradiation. However, vortex depinning is found to be enhanced by photon-induced reduction in potential barrier, which mitigates the adverse effect of film inhomogeneity on superconductivity in the *a*-MoSi thin films. The thorough understanding of the vortex dynamics in *a*-MoSi thin films under the effect of external stimuli is of paramount importance for both further fundamental research in this area and optimization of future superconducting devices.

# 1. Introduction

In type-II superconductors, as introduced by Abrikosov[1], when the external magnetic field slightly exceeds the lower critical field $H_{C1}$, the magnetic flux starts to penetrate through the superconductors. As the Meissner effect is destroyed, there will be whiskers of normal phase surrounded by superconducting phases, with each normal core circulated by supercurrents called Abrikosov vortex (abbreviated as vortex). The magnetic field through each vortex and its neighboring area is quantized with flux $\Phi_0$ of $2.06783 \times 10^{-15}$ Wb, *i.e.*, magnetic quantum flux. When the applied magnetic field is greater than the upper critical field $H_{C2}$, vortex cores overlap and the magnetic field penetrates the entire material, leading to the disappearance of vortices. Between $H_{C1}$ and $H_{C2}$, the vortex will persist with its core made up of unpaired electrons. Vortex motion can be described by classical mechanics due to its particle nature. A vortex also exhibits the nature of waves, giving rise to its temperature dependent probability of tunneling from one point to another. Thus, vortex dynamics, which can be regulated by multiple external stimuli such as electric field, magnetic field, temperature and light, have long been a fascinating facet of superconductors[2-6].

For a current biased superconductor thin film, external stimuli such as light (electromagnetic radiation) could be a trigger for transition from a non-energy dissipating state to a resistive state. An incident photon of energy $h\nu$ will generate a normal domain (*i.e.,* hot spot) where quasiparticle numbers are in the order of $h\nu/\Delta$. In a superconductor biased slightly below its critical current $I_C$, the current will subsequently redistribute and exceed the local critical current density $J_C$ in the proximity of the hot spot, leading to the growth of the hot spot into a hot belt across the superconductor[7]. The formation of such a hot belt across the superconductor, which causes a jump in the voltage, has been exploited as an effective route of responding to electromagnetic radiation, playing a vital role in a number of emerging fields such as quantum computing and deep-space exploration[8-10]. The presence of vortices provides an alternative platform for this transition to take place. Vortex is subjected to the Lorentz force in the presence of a magnetic field. Increasing current makes the Lorentz force exceed the pinning force, which leads to vortex motion, *i.e.*, the movement of unpaired electrons. This creates electric field inside the superconductor, allowing the quasiparticles in vortex core to gain enough energy to escape via Andreev reflection[11]. As a consequence, the size of the vortex core gradually decreases, resulting in a decrease in the viscous coefficient of vortex motion, which increases the vortex movement. When the Lorentz force reaches its maximum, the flux flow will lose its stability, resulting in a jump in current-voltage (*I-V*) curve. Upon irradiation, the induced vortex-antivortex pairs in the hot spot driven by the Lorentz force toward opposite direction can also generate a signal caused by flux flow instability that can be captured by the external circuit[12].

By far, electromagnetic radiation is currently considered more as a trigger, *i.e.*, impulse signal, in the above-mentioned physical processes. However, for type-II superconductors, continuous external electromagnetic radiation must also be taken into account due to the presence of the vortex, as it may have an effect on the different stages of the vortex dynamics. For example, the vortex may absorb photon energy and move at a larger velocity, or reduce the vortex pinning force, both of which make the system more likely to reach the flux flow instability. Therefore, understanding the influence of continuous irradiation on vortex dynamics is of great significance for rational design of future superconducting devices. While amorphous MoSi (*a*-MoSi) has attracted extensive research for superconducting devices because of its small energy gap and simple fabrication process, corresponding to superior performance at longer wavelength and ease of deposition over large areas[13-19], there is still a lack of study on vortex dynamics in *a*-MoSi superconducting thin films. In this work, we have selected a batch of *a*-MoSi films prepared under the same processing conditions to understand their vortex dynamics in a more comprehensive and systematic manner. The entire study is conducted based on a concise way of examining vortex dynamics through measuring the electronic transport of *a*-MoSi thin films in the coexistence of magnetic fields and irradiation[10].

## 2. Methods

*a*-MoSi thin films with thickness of ~20 nm were deposited onto thermal silicon oxide substrates by magnetron sputtering under high vacuum (base pressure < $7.5 \times 10^{-8}$ torr). Detailed fabrication process and characterization of *a*-MoSi thin films can be found elsewhere[20,21]. 5×5 mm *a*-MoSi thin films were bonded in 4-wire configuration with 1 mm interval by ultra-sonic bonding method. All the electrical measurements were done using a physical property measurement system (PPMS, Cryogenic CFMS-14T). The magnetic field is applied in two directions, *i.e.,* perpendicular to the sample surface (out-of-plane, OP) and parallel to the sample surface (in-plane, IP), respectively. The standard four-wire method is used to eliminate the influence of contact resistance on the measurements. A 360 nm diode laser with intensity of ~20 mW·cm$^{-2}$ was chosen as the light source for irradiation.

## 3. Results and discussion

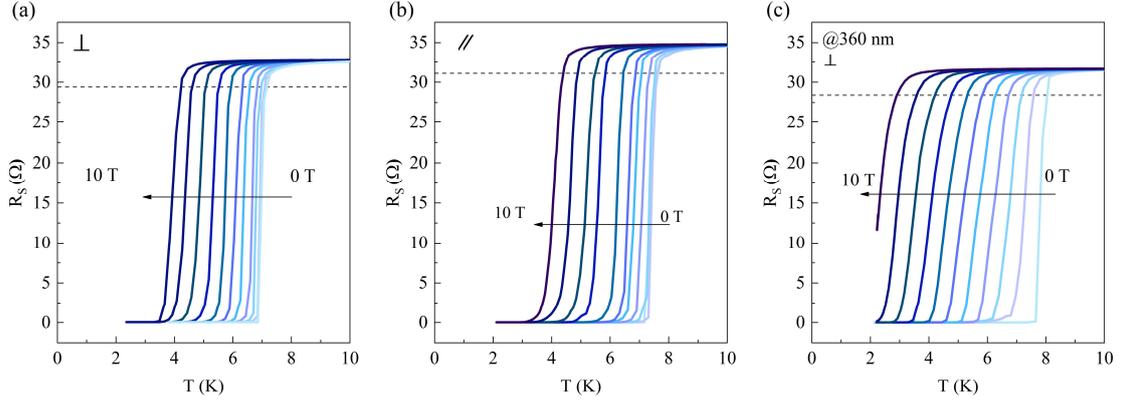

**Figure 1. Sheet resistance $R_S$ of $a$-MoSi thin films as a function of temperature $T$.** (a) under OP magnetic field, (b) under IP magnetic field, (c) under 360 nm laser irradiation and out-of-plane magnetic field.

Among the two critical fields $H_{C1}$ and $H_{C2}$ that define the field span in which vortices are present, the latter is more often reported due to its ease of acquisition. In order to determine $H_{C2}$ of our $a$-MoSi thin films, the temperature dependent sheet resistance $R_S$ under external magnetic field was obtained, as shown in Figure 1. In order to avoid sample damage and waste of time, $a$-MoSi film deposited on one-inch silicon wafer was cut into small pieces of different sizes and loaded onto corresponding sample holders for different tests (IP, OP and illumination). This leads to fluctuations in the critical temperature $T_C$ in the absence of a magnetic field in the range of 7.4 ± 0.4 K, in agreement with previous reports[21,22], where the fluctuation may come from the non-uniform superconducting properties of $a$-MoSi thin films. The variation of $T_C$ with the external field is not affected by this fluctuation, and a tendency to decrease with increasing magnetic field is observed for different magnetic field directions. Additionally, under the same 10 T magnetic field, the superconducting transition curve of the sample exposed to 360 nm laser irradiation has exceeded the lowest temperature of PPMS (2 K) (as shown in Figure 1(c)), while the $T_C$ of the samples not exposed to light are still within the measurement temperature range (> 2 K). It is clear that light causes the $T_C$ to shift towards lower temperatures. Here the $H_{C2}$ are determined by applying the constant 90% $R_S$ criterion (shown by the dash lines)[23], which indicates the onset of the normal state.

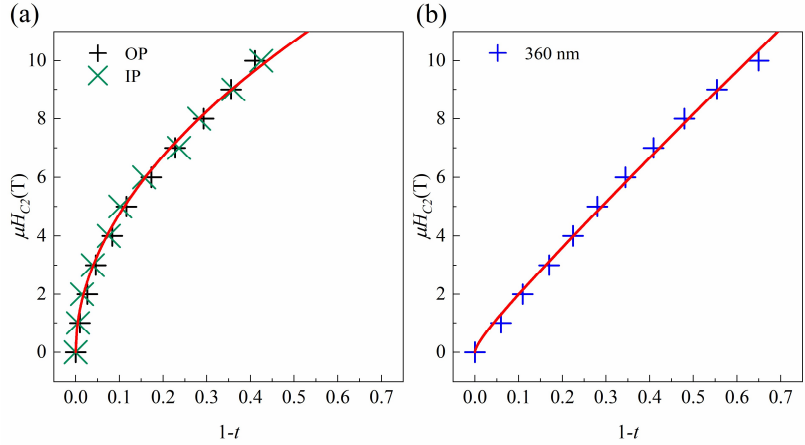

**Figure 2. Temperature dependence of $H_{C2}$.** (a) Under OP and IP fields without light irradiation. (b) under OP field and a 360 nm laser irradiation. Here $t=T/T_C$ is the reduced temperature.

Several insights can be gained from the results of temperature dependent $H_{C2}$ as shown in Figure 2. First, the $H_{C2}$ data obtained in OP and IP directions are nearly identical just as $T_C$. Thus, the following study and analysis will be focused on the scenario that the samples are subjected to OP magnetic field only. Second, it can be seen from Figure 2(a) that the $H_{C2}$ curves bend significantly closer to $T_C$ in the absence of light. It is well known that when $\xi<d$ (film thickness), the $H_{C2}$-T relation could be described by 3D Ginzburg-Landau (GL) model[24]

$$\mu_0 H_{C2} = \frac{\Phi_0}{2\pi\xi(0)^2} \times (1-t), \quad t = \frac{T}{T_C} \quad (1)$$

where $\mu_0$ is the vacuum permeability, $\Phi_0$ is the magnetic quantum flux mentioned before and $\xi(0)$ is the coherence length at T=0 K. When $\xi>>d$, the 3D GL model does not apply and is replaced by a 2D GL model proposed by Harper and Tinkham[25]

$$\mu_0 H_{C2} = \frac{\sqrt{3}\Phi_0}{\pi d\xi(0)} \times (1-t)^{\frac{1}{2}}, \quad t = \frac{T}{T_C} \quad (2)$$

So, if the film is at a specific thickness $d$, as the temperature approaches $T_C$, the cooper pairs gradually expand and $\xi$ will experience the transition from $\xi<d$ to $\xi>>d$. Consequently, the linearity of the $H_{C2}$ curve at low temperatures will disappear near $T_C$, as shown in Figure 2(a). Similar 2D-3D crossover was found in other studies[26-30]. The fitting lines here are a combination of Equation (1) and (2), covering both 2D and 3D cases. It yields a $\xi(0)$ of ~5 nm, suggesting our *a*-MoSi thin films are in dirty limit i.e., $\xi>>l$ when compared with the mean free length $l$ ($\approx 0.2$ nm) reported in previous studies[22]. Additionally, such non-linearity may also be ascribed to the non-uniformity of the films that will lead to suppression of pair-breaking[31]. The localization of the superconducting order parameter caused by random superconducting nanomesh will quench the orbital pair breaking of the OP field. However, such non-linearity near $T_C$ largely diminishes and the experimental data shows a good linearity when the sample is irradiated by a 360 nm laser, as shown in the Figure 2(b). This may be due to the suppression of the cooper pair expansion by light and is subject to further study.

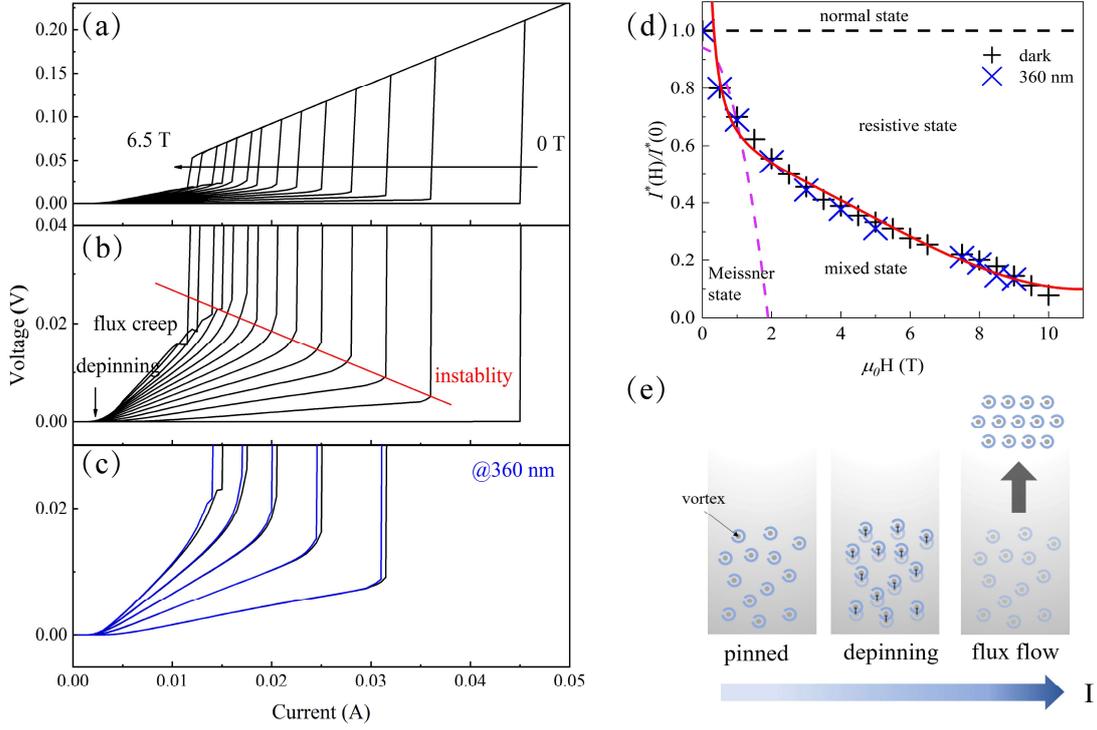

**Figure 3. Vortex dynamics behaviors.** (a) *I-V* curves of the *a*-MoSi thin film at *T*=2 K under different external magnetic fields and (b) corresponding enlarge graph, (c) under 360 nm laser irradiation, (d) magnetic field dependent critical current $I^*$, (e) schematic diagram of vortex dynamics in our *a*-MoSi thin films.

The *I-V* curves of our *a*-MoSi thin films in Figure 3(a) implies the evolution of the vortices states under the effect of continuously varying Lorentz force. Here we mark the critical points ($I^*, V^*$) on the *I-V* curve as the onset of normal state (shown by the red line in Figure3(b)). In the Meissner state (vortex free), $I^*$ indicates transition that is unrelated to vortex dynamics, as shown by the abrupt first order transition of the 0 T curve in Figure 3(a). The dependence of $I^*$ on magnetic field could be expressed as[32]

$$I^*(H) = i_s w(1 - 6h^2 + 3\sqrt{3}h^3), \quad h = \frac{\pi \xi w H}{\Phi_0} \qquad (3)$$

where $i_s$ is the pair-breaking current density and $w$ is the film width. In contrast, in the mixed state $I^*$ indicates the onset of normal state caused by flux flow instability, exhibiting a power law with the magnetic field[32,33]

$$I^*(H) = i_p w + \frac{A_C^2}{4H\alpha} \left[ \frac{5}{9} - \frac{4i_p \alpha}{A_C} \left( q_p + \frac{1}{\sqrt{3}} \right) - 2q_p^2 + q_p^4 \right], \quad (4)$$

where $i_p$ is the depinning current density, $q_p$ is a dimensionless vector potential measured in units of $A_C = \Phi_0/2\pi\xi$ corresponding to $i_p$ and $\alpha = 2\pi\lambda/c$. As shown in Figure 3(d), the data was well fitted by the combination of Equation (3) and (4) (shown by the red line). The purple dotted line here is the dependence of the critical current on the magnetic field in the Meissner state, below which the material is in a vortex-free state. Above the critical current (red line), there is a vortex flow inside the material which corresponding to the resistive state. Between the Meissner state and resistive state, the material is filled with pinned vortices, *i.e.*, a mixed state. When the Lorentz force ($F_L$) exerted on the vortices is weaker than its pinning force ($F_L<F_P$), the vortices remain localized to where the defect resides, as illustrated by the left diagram of Figure 3(e). Here the direction of the bottom arrow represents the direction of the current and the shade of the blue represents the magnitude of the current. As the $F_L$ increases, the vortices start to creep from their pinning sites ($F_L \approx F_P$). This depinning process is sketched in the middle diagram of Figure 3(e). When the $F_L$ is sufficiently strong, the vortices will move freely, driven by the Lorentz force ($F_L>F_P$), to generate a flux flow through the superconductor (shown in the right diagram of Figure 3(e)). Such a continuous process is highly consistent with the Landau second order phase transition and is in stark contrast to the abrupt first order transition in the absence of a magnetic field (0 T curve in Figure 3(a)). Under a 360 nm laser irradiation, $I^*$ slightly declines while $V^*$ remains almost unchanged as shown in Figure 3(c).

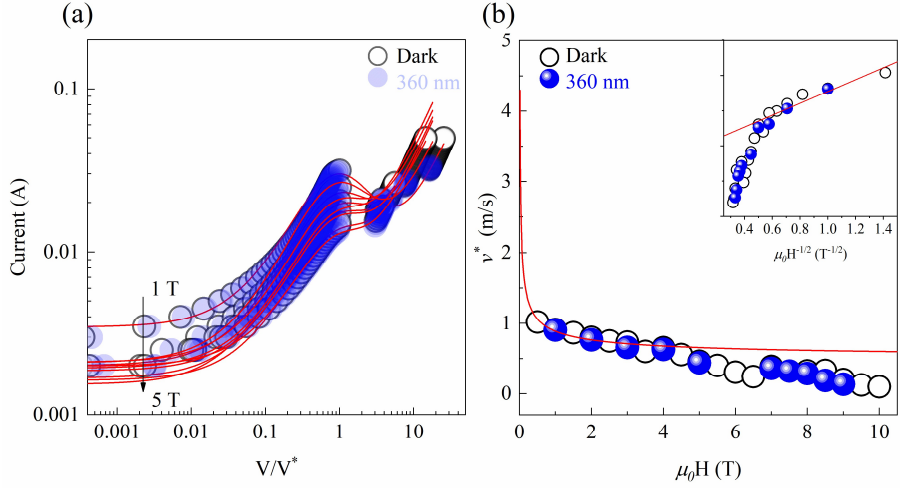

**Figure 4. LO critical velocity $v^*$ and quasiparticle inelastic scattering time $\tau^*$.** (a) LO instability fits to experimental V-I curves (b) magnetic field dependent critical velocity $v^*$.

When a vortex acquires energy $\delta E$ from the external field, its unpaired electrons will turn into hot electrons similar to those release from the destruction of Cooper pairs by high energy photons. Based on Larkin-Ovchinnikov (LO) theory[11,34], the quasiparticle inelastic scattering will lead to a reduction in the viscous damping coefficient $\eta(v)$

$$\eta(v) = \eta(0) - \eta(0)\delta E/\Delta \qquad (5)$$

where $\eta(0)$ is the viscous damping coefficient at zero velocity and $\Delta$ is the energy gap of superconductor. When the energy of the quasiparticle is in dynamics equilibrium, i.e., $\frac{jE}{n} = \delta E/\tau^*$ with $\tau^*$ is the inelastic scattering time, the viscous damping coefficient $\eta(v)$ could be expressed as

$$\eta(v) = \frac{\eta(0)}{1+(v/v^*)^2} \quad (6)$$

Obviously, as the vortices move faster, the $\eta(v)$ decreases so that the material transforms into the unstable flux-flow state at the critical velocity $v^*$

$$v^{*2} = \frac{1.31D}{\tau^*}\sqrt{1-T/T_C} \quad (7)$$

where $D=v_f l$ is the quasiparticle diffusion constant. From the LO theory, vortex instability reflected by the *I-V* curves can be described by the phenomenological equation[9]

$$I(V) = \frac{V}{R_S}\left[\frac{\alpha}{1+\left(\frac{V}{V^*}\right)^2} + \frac{\beta\left(\frac{V}{V^*}\right)^{-c}}{1+\left(\frac{V}{V^*}\right)^2} + 1\right] \quad (8)$$

where $\alpha$, $\beta$ and $c$ are fitting parameters that depend on the strength of magnetic field. As shown in Figure 4(a), the LO vortex instability (shown by red lines) fits well with our experimental data. The $v^*$ could be obtained through $v^*=E^*/B=V^*/dB$, where $d$ is distance between two electrodes, as shown in Figure 4(b). Under small magnetic field, the mean free length $l$ is shorter than the intervortex length $a$, which means the quasiparticle distribution has a strong partial spatial inhomogeneity, mixed with a superconducting phase with equilibrium quasiparticle particle distribution. When the quasiparticles are moving at a critical velocity $v^*$, $l=v^*\tau^*=a$ and can be expressed as[35]

$$v^* = C\left(\frac{\Phi_0}{\mu_0 H}\right)^{\frac{1}{2}}\frac{f(T)}{\tau^*} \quad (9)$$

where *C* is a constant relating to the geometry of vortex lattice and *f(T)* is a numerical factor of order 1. The experimental data at low fields (<4 T) are consistent with the expectations of the theoretical model, as shown by the red lines in Figure 4(b). When the magnetic field increases further, $l$ will become larger than $a$ and the quasiparticles will distribute uniformly through the whole system, leading to $v^*$ being independent of the magnetic field. In our case, however, $v^*$ tends to zero under high magnetic fields, similar to the results found by Hofer et al. in nanocrystalline $\beta$-$W$ thin films[36]. Notably, magnetic field-independent $v^*$ is observed in superconducting nano/micron wires[9,10], while in larger sized superconducting films it is replaced by a $v^*\rightarrow 0$ behaviour at high fields. This may be ascribed to the different velocity distributions of the two cases. In our case, the velocity distribution is broadened by a uniform vortex flow over the entire area between the two electrodes, with either fast or slow velocities over different areas, resulting in experimental value (average value) that converges to zero at high fields. However, the vortex velocity distribution of superconductors that made narrow enough or doped to guide the vortex motion through a narrow channel can be sharpened[37], resulting in a large experimental value of $v^*$. Further research is needed on the specific mechanisms of this $v^*\rightarrow 0$ behaviour. It can be observed that light irradiation does not induce a significant change in vortex velocity, but merely lower the Lorentz force required for triggering the flux-flow state. Hence, light irradiation may reduce the activation energy required for the departure of vortices from the pinning sites.

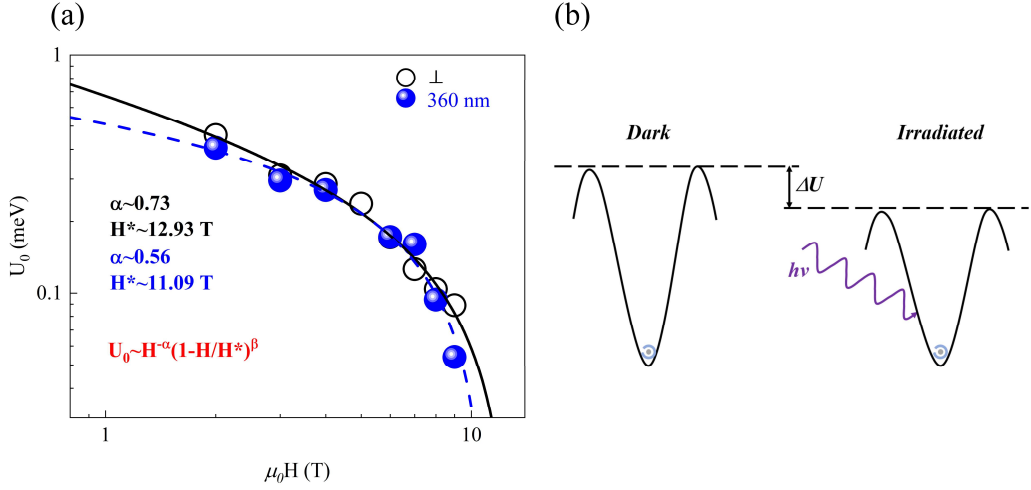

**Figure 5. Thermal activation of vortex.** (a) magnetic field dependent activation energy $U_0$, (b) schematic diagram of vortex tunneling.

As discussed above, a vortex pinned at one point has a finite probability of tunneling through to another pinning site. The variation in this microscopic tunneling probability will lead to a change in the macroscopic resistance $R(T)$ of the material, its relationship could be formulated as the Arrhenius function[23]

$$R(T) = R_0 \exp\left(-\frac{U_0}{k_B T}\right) \quad (10)$$

where $U_0$ is the activation energy of the vortex, $R_0$ is the pre-factor. By plugging the sheet resistance Rs (Figure 1) into Equation (6), the dependence of $U_0$ on the applied magnetic field $H$ is worked out as shown in Figure 5(a). The $U_0$ - $H$ relation could be well described by[38]

$$U_0 \propto H^{-\alpha}(1 - H/H^*)^\beta \quad (11)$$

where $\alpha$ and $\beta$ are the fitting parameters, $H^*$ is the irreversible field at $T=0$ K that demarcates the vortex glass from vortex liquid. A slight reduction in $U_0$ can be clearly seen under light irradiation. This is because that light provides energy for the vortex to displace from the pinning point, leading to a decrease in the effective potential barrier as shown in Figure 5(b). As the process is not quantized, the range of wavelengths of light used to promote vortex motion can be moved closer to the infrared end. When the width of the superconductor is sufficiently small, the mid-infrared signal is able to trigger vortices across the superconductor and thus produce a response, as reported by Chen et al.[20].

## 4. Conclusion

In summary, we studied the continuous irradiation effect on vortex dynamics in superconducting $a$-MoSi thin films through electrical transport measurements under external magnetic fields. $a$-MoSi thin films exhibit Landau second order phase transition under external magnetic fields with the coherence length $\xi(0)$ of ~5 nm at $T=0$ K. The quasiparticle critical velocity $v^*$, which was determined from the flux flow instability, shows a power law dependence on magnetic field in the low field range ($v^* \sim B^{-1/2}$) and tends to zero under strong fields. The vortex depinning is promoted by light irradiation through lowering the effective potential barrier, thereby facilitating the vortex motion. Such photon-enhanced vortex depinning will alleviate the non-uniformity of the film, resulting in a more homogenous superconductivity. Our work fills the knowledge gap in vortex dynamics of $a$-MoSi thin films by demonstrating the effect of light irradiation on the vortex depinning, which confirms that the optoelectronic performance of amorphous superconductors is not seriously affected by the quality of films, consolidating their competitive advantages for applications in single-photon detectors.

**Acknowledgments**

This work is partially supported by National Key Research and Development Program of China (2017YFB0503300), the Fundamental Research Funds for the Central University (No: 310201911cx024) and Natural Science Foundation of Shaanxi Province (No: 2021JM-059). We also want to thank Analytical & Testing Center of Northwestern Polytechnical University for the help for electrical tests.